\documentclass[structabstrac]{aa}
\usepackage{txfonts}
\usepackage{natbib}
\usepackage{rotating}
\usepackage{subfigure}
\usepackage{multirow}
\usepackage{color}

\begin{document}


\title{The Quasi-Biennial Periodicity (QBP) in velocity and intensity helioseismic observations}
\subtitle{The seismic QBP over solar cycle 23}


\author{R.~Simoniello\inst{1} 
\and W.~Finsterle\inst{1}
\and D.~Salabert\inst{2}
\and R.~A.~Garc\'ia\inst{3}
\and S.~Turck-Chi\`eze\inst{3}
\and A.~Jim\'enez\inst{4}
\and M.~Roth\inst{5}}
\institute{PMOD/WRC Physikalisch-Meteorologisches Observatorium Davos-World Radiation Center, 7260 Davos Dorf, Switzerland 
\email{rosaria.simoniello@pmodwrc.ch,wolfgang.finsterle@pmodwrc.ch}
\and Universit\'e de Nice Sophia-Antipolis, CNRS UMR 6202, Observatoire de la Cote d' Azur, BP 4229, 06304 Nice Cedex 4, France 
\email{salabert.david@oca.fr}
\and Laboratoire AIM, CEA/DSM-CNRS-Universit\'e Paris Diderot; CEA, IRFU, SAp, centre de Saclay, F-91191, Gif-sur-Yvette, France
\email{rgarcia@cea.fr,sylvaine.turck-chieze@cea.fr}
\and IAC, Instituto de Astrofis\'\i ca de Canarias, 38205, La Laguna, Tenerife, Spain
\email{ajm@iac.es}
\and Kiepenheuer Institute for Solar Physics, Freiburg 
\email{mroth@kis.uni-freiburg.de}}
\date{Received 9 September 2011/ Accepted 12 January 2012}
\abstract{}{We looked for signatures of Quasi-Biennial Periodicity (QBP) over different phases of solar cycle by means of acoustic modes of oscillation. Low-degree $p$-mode frequencies are shown to be sensitive to changes in magnetic activity due to the global dynamo. Recently have been reported evidences in favor of two-year variations in $p$-mode frequencies.}
{Long high-quality helioseismic data are provided by BiSON ({\bf Bi}rmingham {\bf S}olar {\bf O}scillation {\bf N}etwork), GONG ({\bf G}lobal {\bf O}scillation {\bf N}etwork {\bf G}roup), GOLF ({\bf G}lobal {\bf O}scillation at {\bf L}ow {\bf F}requency) and VIRGO ({\bf V}ariability of Solar {\bf IR}radiance and {\bf G}ravity {\bf O}scillation) instruments. 
We determined the solar cycle changes in $p$-mode frequencies for spherical degree $\ell$=0, 1, 2 with their azimuthal components in the frequency range 2.5~mHz~$\leq~\nu~\leq$~3.5~mHz. }
{We found signatures of QBP at all levels of solar activity in the modes more sensitive to higher latitudes. The signal strength increases with latitude and the equatorial component seems also to be modulated by the 11-year envelope.}
{The persistent nature of the seismic QBP is not observed in the surface activity indices, where mid-term variations are found only time to time and mainly over periods of high activity. This feature together with the latitudinal dependence provides more evidences in favor of a mechanism almost independent and different from the one that brings up to the surface the active regions. Therefore, these findings can be used to provide more constraints on dynamo models that consider a further cyclic component on top of the 11-year cycle.}
\keywords{Methods: data analysis - Sun: helioseismology - Sun: activity}
\titlerunning{QBP from helioseismic observations}
\maketitle
\section{Introduction} 
{Evidences in favor of shorter periodicities than the 11-year period, ranging from days to years in solar activity proxies, are very well known in literature. For short-terms, the Sun often exhibits 27-day and 13.5-day periodicities, while the regime between days and 11 years are referred as mid-term periodicities \citep{Bai03}. Among these, particularly prominent are the ones between 1.5-4 years \citep{Vec09}. We will refer to them as Quasi-Biennial Periodicity (QBP). This phenomenon has been observed in radio flux on 10.7cm \citep{Bel96}, total solar irradiance \citep{Pen06}, flares and sunspots \citep{Aki87, Mur03, Val08} and green coronal emission line \citep{Vec08,Vec09}. It also seems to appear only time to time tending to be more important over periods coinciding with solar maxima and are not recognized as typical feature of every cycle \citep{Kri02,Kna04,Cad05,Li06}. This suggests that the QBP might be closely linked to the strength of the global dynamo.}
 Coronal data have shown a latitudinal dependence of the signal strength \citep{Vec08}. It is particularly strong in the polar region and weaker around the equator and more over the periodicity from northern and equatorial regions ($\approx$2.8 years) is different from the one in southern regions ($\approx$1.5 years). More recently the same authors, investigating a longer dataset recorded from 1939 to 2005, concluded that the non-constant period lengths is the manifestation of a unique quasi-biennial activity cycle \citep{Vec09}.

Since the discovery of the correlation between the observed global acoustic mode ($p$-mode) parameters and the solar magnetic activity cycle \citep[e.g.][]{Woo85,Pal89,Els93}, it became evident that the acoustic modes can also be used as  a diagnostic tool of the solar cycle. 
In particular with high-quality and continuous helioseismic data, deeper analysis revealed that mode amplitudes are suppressed with increasing magnetic activity, while mode frequencies are shifted towards higher values as the solar cycle approaches its maximum at low and high-angular degree
\citep{How99,Kom00,Jim01,Cha01, Sal04, Sal06, Sim10}. While for the mode amplitudes a possible explanation in terms of mode conversion has been provided \citep{Sim10}, we still lack the basic understanding of the mode frequency shift. $F$-modes can help to understand this variability and the confrontation of observations to predictions seem to show some structural effects \citep{Syl09}. Unfortunately, a proper introduction of sub-surface magnetic field is still lacking. So this does not allow to properly estimate the structural consequences of this important factor. The form and the degree dependence of the shifts, anyway favor near-subsurface phenomenon, but however they cannot be purely explained by structural changes, the magnetic field is somehow involved in the mechanism \citep{Goo02, Dzi04}. Recently the argument that the solar cycle variation in the global mode frequency are due to the global averaging of active regions \citep{Hin01} has been only partially supported by the observations \citep{Tri10}. It has been argued that the weak component of the magnetic field (turbulent or horizontal field) must be taken into account in order to fully explain the observed mode frequency shift. 

Recent observations in $p$-mode frequencies
   have also shown the presence of a 2-year modulation most prominent over periods coinciding with solar maxima. The properties of the seismic QBP have been interpreted as the visible manifestation of a second dynamo mechanism \citep{Ben95,Ben98a,Ben98b}, induced by the strong latitudinal shear located below the surface at $\approx$ 0.95 solar radius \citep{Schou98}. The second dynamo itself cannot explain the observed modulation in the signal strength and therefore it has been supposed that
 when the 11-year cycle is in its strong phase, buoyant magnetic flux are sent upward by the main dynamo into shallower layers, thus allowing the seismic QBP to be mainly detected over periods of solar maxima \citep{Fle10}.  
However it is still a matter of debate if this second dynamo mechanism is indeed behind the QBP observed in activity proxies and in solar seismic data. In fact,
recent observations of rapidly rotator stars have shown signatures of QBP, whose strength increases with increasing magnetic activity and it gets stronger at polar regions. It has been supposed that the unstable magnetic Rossby waves in the tachocline might be behind it \citep{Zaq10,Zaq11}. If this is also true for solar type star is still an undergoing investigation.

 The $p$-mode spatial configuration can be described by spherical harmonics with each mode being characterized by its spherical harmonic degree $\ell$ and azimuthal order $m$. Modes with $\ell \leq$ 3 are named low-degree modes and conversely to surface activity indices, they also sound the interior of the Sun being shaped in deeper layers. Therefore they might feel magnetic changes that have yet to manifest at the surface or it might also be possible that the disturbances will never reach the Sun's surface. 
 By averaging the individual $p$-mode frequency shifts over all low ($\ell$)- values, observational findings pointed to seismic signatures of QBP over the long extended minimum characterizing the end of solar cycle 23 and beginning of solar cycle 24 \citep{Bro09a,Fle10}. Further evidences in favor of this feature have been reported in low-degree modes by investigating the spherical degree dependence \citep{Bro11}, but some authors have cast doubts, because they didn't spot any significant trend during periods of low activity \citep{Jai11}. We decided, therefore, to investigate the properties of the seismic QBP by using several helioseismic instruments in velocity and intensity data covering a period that goes from the end of solar cycle 22 up to the beginning of solar cycle 24. Each observational program can provide more stable and/or accurate data depending on instrumental features \citep{Bro09b,Sal11}. Therefore by comparing different observations we have been able to check whether or not the seismic QBP occurs only time to time and during periods of high activity. Furthermore
we carried out this analysis for each of the sectoral ($m$=$\ell$) and zonal ($m$=0) components of low-degree modes (without averaging over the azimuthal components) and this is the novelty of the work within this topic.
We also investigated if and to which extent the mid-term periodicity is linked to the appearance of active regions induced by the main dynamo. To this aim we carried out a correlation analysis between each azimuthal ($m$)-components of low-degree mode frequencies and activity indices linked to strong and/or weak magnetic fields over different phases of solar cycle 23. It is important to investigate the phasewise 
variation in correlation between low-$\ell$ frequencies and 
activity indices as well, because earlier study of the $m$-components were confined to intermediate degree modes only \citep{Jai09}. Furthermore, acoustic waves, depending on their spherical degree and azimuthal component, are more sensitive to low-mid latitudes or higher latitudes. Therefore the analysis of the azimuthal components allowed us also to have some hints on the role played by strong and/or weak fields with latitudes as sources of the QBP.
 
The seismic investigation of the mid-term periodicity might help us to gain a deeper insight on the global properties of the Sun as for example on the solar dynamo theory as well as on stellar cycles: mid-term periodicities have already been observed in fast rotator stars and therefore the better understanding of this short periodic behavior in the Sun might help illuminate the interpretation of this signal of solar-like stars.
\section{Helioseismic data analysis}
\subsection{Mode visibility}
{Low-degree modes are detected either in Sun-as-a-star observations or through highly spatially resolved images. In the first scenario the modes are observed in averages made over the visible disk of Doppler velocity or intensity variations.  
In this case only the modes that satisfy the condition $\ell+m=even$ are visible, because of the near-perpendicular inclination of the Sun with respect to the Earth. In observations with high spatial resolution, the visible disk is sampled in many pixels and the collected images are decomposed into their constituent spherical harmonics, giving access to all 2$\ell$+1 component for each degree.}
\subsection{Sun-as-a-star observations from ground based network}
{\bf Bi}rmingham {\bf S}olar {\bf O}scillation {\bf N}etwork (BiSON) has been operating since 1976 and makes unresolved solar disc observations. It consists of 6 resonant scattering spectrometers, that perform Doppler velocity measurements in integrated sun light of the K Fraunhofer line at 7699\AA~\citep{cha95}. The data are dominated by the Doppler variations from the low-degree, $\ell$, modes.
The nominal height is at $\approx$260~km above the photosphere \citep[e.g.][]{jim07}. {We analyzed the data provided by BiSON instruments from 11th April 1996 to 4th of April 2009. This long dataset was split into contiguous 365-day  series with four overlaps of 91.25 days.
Estimates of the $p$-mode frequencies for $\ell$=0,1,2 were extracted from each subset by fitting a modified Lorentzian model to the data using a standard likelihood maximization method \citep{Fle09}.}
\subsection{Sun-as-a-star observations from space}
The {\bf G}lobal {\bf O}scillation at {\bf L}ow {\bf F}requency (GOLF) instrument, on board the {\bf SO}lar and {\bf H}eliospheric {\bf O}bservatory (SOHO) satellite, is devoted to the search of low-degree modes (Gabriel et al. 1995). It works by measuring the Doppler shifts of the Na line at 5889\AA~(D1) and 5896\AA~(D2). Due to the malfunctioning of the polarization system, GOLF is working in a single-wing configuration and it has been proved that it is almost a pure velocity signal \citep{Pal99}. The one-wing configuration has been changing during the 14 years of observations as follows: 1) 1996-1998 in the blue-wing configuration; 2) 1998-2002 in the red-wing configuration; 3) up today in the blue-wing configuration \citep{Ulr00,Gar05}. 

Due to the formation heights of the spectral lines, in the blue-wing configuration, GOLF observes at $\approx$330~km, while in the red-wing configuration at $\approx$480~km \citep{jim07}. Although the changing observational height affects the $p$-mode amplitude determination \citep{Sim09,Sim10}, it doesn't affect the $p$-mode frequency. {We analyzed the data provided by GOLF instruments from 11th April 1996 to 4th of April 2009 and the estimates of $p$-mode frequency shifts have been extracted by applying the same fitting method as in BiSON. This differs from the one applied to VIRGO and GONG $\ell$=0 integrated. We, therefore, verified that both fittings methods returned consistent results.}

The {\bf S}olar {\bf P}hoto{\bf M}eters (SPM) of the {\bf V}ariability of solar {\bf IR}radiance and {\bf G}ravity 
{\bf O}scillations (VIRGO) package \citep{Fro95,Fro97} aboard SOHO satellite, consists of three independent 
photometers, centered around 402, 500 and 862 nm (the blue, green, and red channel 
respectively). They measure the spatially integrated solar intensity over a 5 nm
bandpass 
at a one minute cadence. The data obtained by the VIRGO/SPM instrument over the 
mission have been of uniformly high quality, whether before or after the loss 
of contact with the spacecraft for several months in 1998. Each one of the three
different
signals observes the Sun-as-a-star oscillations at different heights in the solar atmosphere \citep{Jima05}.

{The data provided by VIRGO instruments cover almost 14 years 
of data, starting on the 11th April 1996 up to 8th April 2010.
 This dataset was split into contiguous 365-day with four overlaps of 91.25 days. The power spectrum of each subseries was fitted to estimate the mode $p$-mode parameters for $\ell$=0,1,2 using a standard likelihood maximization function  \citep{Sal07}. Each mode component is parameterized using a asymmetric Lorentzian profile.}
\subsection{High-resolved observations from ground based network}
The {\bf G}lobal {\bf O}scillation {\bf N}etwork {\bf G}roup (GONG) consist of six instruments deployed worldwide to provide nearly continuous and stable velocity images of the Sun. 
GONG instrument is based on a Michelson interferometer called a Fourier Tachometer \citep{Bec78}. It works by using the Ni line at 676.8nm and the observational height is approximately at 213~km. The network started to be fully operating since May 1995. {Time series of 36 days (the so-called GONG month) are produced and the 
individual frequencies of the azimuthal components are determined  
up to $\ell=150$ over subseries of 108 days by 
concatenating three GONG months. These frequencies are afterwards made 
publicly available (http://gong2.nso.edu/archive/).} 

{The $\ell$=0 GONG time series between May 7th 1995 and May 3rd 
2008 was analyzed in a similar manner as for VIRGO data, returning 
estimates of the mode frequencies of $\ell$=0,1,2. In order to analyze the azimuthal components individually, we made use of the frequency tables of the $\ell$= 1,2 modes available from the GONG webpage between the 29th of June 
1995 and the 24th of December 2009. Very few outliers were removed from the 
following analysis using the quality flags provided by the GONG team.}

\subsection{Frequency shift determination}
{The temporal variations of the $p$-mode frequencies were defined as the 
difference between the frequencies of the corresponding modes observed 
at different dates and reference values taken as the average over the 
years 1996-1997. The weighted averages of these frequency shifts were 
then calculated between 2500 and 3500~$\mu$Hz. 
Furthermore for each spherical degree the shifts from the VIRGO green, blue and red channels have been averaged. The shifts for each spherical degree with the same $\mid m \mid$ obtained by the azimuthal analysis of GONG high-resolved data have also been averaged.}
\section{QBP signatures in $p$-mode frequencies over solar cycle 23}
\subsection{Spherical degree dependence of the QBP over the 11-year cycle}
\begin{figure}
\begin{center}
\includegraphics[width=3.2in]{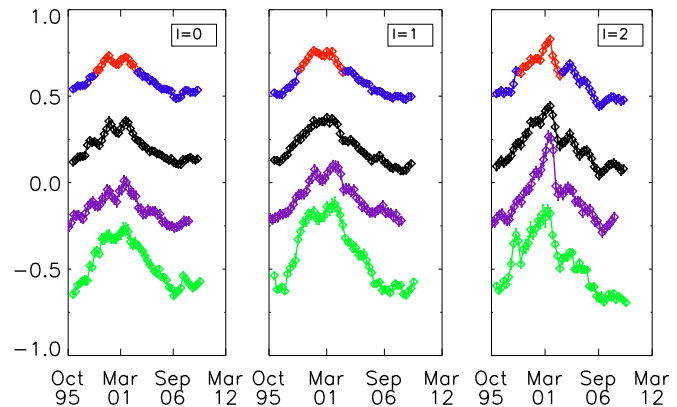}
\caption{Solar cycle changes in $p$-mode frequency shifts ($\mu$Hz) for $\ell$=0, 1, 2 from VIRGO (green), BiSON (black), GONG (purple) and GOLF (blue and red to identify the corresponding wing configuration) observations over solar cycle 23 and beginning of solar cycle 24.} \label{fig:bisongongvirgolf}
\end{center}
\end{figure}

{We looked for the existence of QBP signatures in $p$-mode frequency shifts over the 11-year cycle. Fig.~\ref{fig:bisongongvirgolf} shows the spherical degree dependence of the shift as seen from VIRGO, BiSON, GONG and GOLF datasets. We introduced an artificial offset between the curves to better compare the seismic features.}
We can spot several seismic signatures that might be linked to the QBP signal:

 - for $\ell$=0 modes over the minimum of solar cycle 22 and beginning of solar cycle 23 (1995-1997), GONG observations show a seismic signature of a QBP. Following the ascending phase of solar cycle 23 in BiSON and GONG, between 1997-1999, we find a further modulation of about 2 years. This signature doesn't appear in GOLF and VIRGO data, probably due to the loss of SOHO in June 1998. 
During the maximum of solar activity the modes show up with a clear double peak structure in all datasets. Over the descending phase BiSON, GOLF and VIRGO agree quite well, showing the same downward trend not characterized by any modulation. GONG data, instead, shows an oscillatory signal around December 2003. 
Towards the end of solar cycle 23 and beginning of solar cycle 24 (2006-2009), we observe a QBP in all dataset although in GOLF and VIRGO data are particularly prominent;  

- in $\ell$=1 modes there are no traces of QBP over the ascending phase in BiSON, GOLF and VIRGO data. Instead GONG shows a modulation around July 1998. Over the maximum phase we found a clear double peak structure in all dataset, but in GONG is more pronounced. In GONG, BiSON and GOLF, over the descending phase, we might spot two further modulations before December 2003 and before September 2006. that are instead absent in VIRGO data. Over the long extended minimum of solar cycle 23 only VIRGO shows a further periodicity after September 2006;

- in $\ell$=2 modes we found over the minimum of solar cycle 22 and beginning of solar cycle 23 (1995-1997) a seismic signature of a QBP from GONG observations. We found the same signature in GONG data in $\ell$=0 mode. Over the ascending phase, a strong peak of almost 2-year length around 1998 in VIRGO and GOLF data is present. This signal might be an artefact due to the loss of the satellite between June and September 1998. Over the maximum phase we found a double peak structure in BiSON, GONG and GOLF observations. The descending phase is characterized by a very well modulated signal of about 2 years, although in BiSON, GOLF and VIRGO is more pronounced. During the minimum of solar cycle 23, we spot in BiSON and GOLF data a clear signature of a further periodicity starting in September 2006 and ending before June 2009.

{In summary we found signatures of biennial modulation in all observational programme, but there are differences in the data sets. The signal strength of mid-term variations is weaker compared to the 11-year signal. In a recent paper the mode visibility in GOLF data appears different in the two wing-configurations, being slightly better during red-wing period \citep{Sal11}. This might explain why the faint signal of the QBP is less prominent in GOLF data.  
For intensity observations further aspects need to be taken into account, like the selection of the spectral region in the observations that affect the mode visibility of the modes in photometric measurements \citep{Nut08}. Furthermore the continuum intensity data samples a different (i.e.deeper) region where the temperature variations induced by the acoustic oscillations may be quite small.}

{By analyzing the spherical degree dependence of $p$-mode frequency shift, we detected likely signatures of the 2-year periodicity at all levels of solar activity, instead surface activity indices mainly show it over periods coinciding with solar maxima. We also noticed that away from the solar maximum, the QBP shows up in each low degree mode over different periods of solar activity phase. This results from the different spatial configuration of the modes over the solar disk. For example, we found seismic signatures over the low activity phase in the modes $\ell$=0, 2 but not in the dipole mode $\ell$=1. In integrated sunlight observations due to the mode visibility, the observed $\ell$=1 frequency is a weighted measurement of the visible components ( $\ell$=1, $|m|$=1), while in the case of the $\ell$=2 mode, the fitted frequency corresponds to the weighted measurement of the zonal ($\ell$=2, $m$=0) and sectoral ( $\ell$=2, $|m|$=2) components \citep{Jim04b}. The spatial structure of the oscillation means that the sectoral $|m|$=$\ell$ are more concentrated around equatorial latitudes than the zonal components ($m$=0), and is particularly true with increasing $\ell$ \citep{Chr02}. As consequence the contribution to the QBP signal over the minimum activity might have its origin at higher latitudes \citep{Sal09}. Furthermore the presence of the QBP signal over periods of low activity is not peculiar of solar cycle 23, as guessed by some authors \citep{Fle10}.}
\subsection{Azimuthal dependence of the QBP over the 11-year cycle}
\begin{figure}
\begin{flushleft}
\includegraphics[width=4.in]{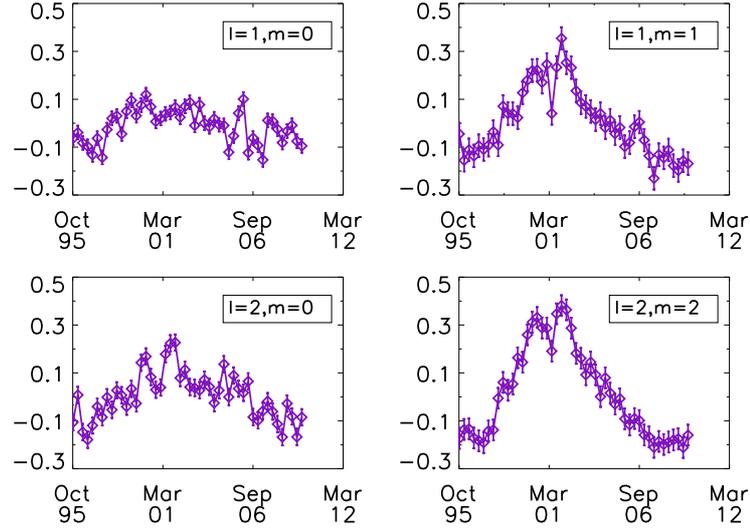}
\caption{Solar cycle changes in $p$-mode frequency shifts ($\mu$Hz) for the $m$-components of $\ell$=1, 2 modes as seen from GONG observations.} \label{fig:mcomp}
\end{flushleft}
\end{figure}

{We used GONG data to look for QBP signal in the azimuthal components of the dipole ($\ell$=1) and  quadrupole ($\ell$=2) mode. Fig.~\ref{fig:mcomp} shows the solar cycle changes in $p$-mode frequency shifts. The sectoral components exhibit a stronger 11-year envelope compared to the zonal ones. This finding clearly confirms the different sensitivity of the modes to magnetic changes with latitude.}
{The $\ell$=1, $m$=0 seems to show a mid-term periodicity over the end of solar cycle 23 with an upward trend starting soon after September 2006. If we look, instead, at $\ell$=1 $m$=1 component, we spot the double peak structure over the maximum of solar activity as also observed in integrated sunlight observations. A further mid-term periodicity seems to occur across September 2006, and the same structure appears in GONG (integrated) and BiSON data. Looking at both $m$-components of the quadrupole mode, the solar maxima is also characterized by the typical double peak structure. Then over the descending phase, the $m$=2 component shows one smoother signature of QBP across September 2006 as in $\ell$=1, $m$=1. The zonal component, instead, seems to show a further periodicity starting soon after September 2006 as we already found in the mode $\ell$=1, $m$=0. Therefore the azimuthal analysis brings more evidences that over the minimum activity phase of solar cycle 23 the QBP might have occurred at high latitudes.}
\section{Latitudinal dependence of the QBP signal strength}
\subsection{Modulation in the QBP signal strength}
\begin{figure}
\begin{center}
\includegraphics[width=3.6in]{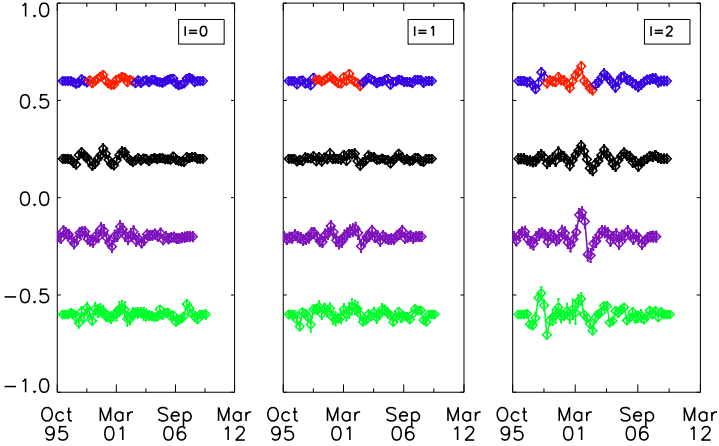}
\caption{The QBP signal in each low-degree $p$-mode frequency shifts after the dominant 11-yr signal has been removed. The color legend is the same as Fig.~\ref{fig:bisongongvirgolf}.} \label{fig:qbo_bisongongvirgolf}
\end{center}
\end{figure}

\begin{figure}
\begin{center}
\includegraphics[width=3.2in]{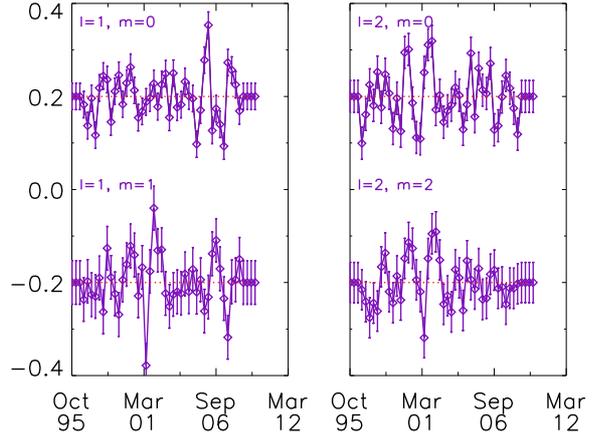}
\caption{The QBP signal in the zonal and sectoral component of the dipole and quadrupole mode frequency shifts after the dominant 11-yr signal has been removed from GONG high-resolved observations.} \label{fig:qbo_gong_l12mcomp}
\end{center}
\end{figure}

{To better reveal the seismic signatures of the QBP signal, we took out the 11-year dependence by using a boxcar of 2.5 years, as already done in \citep{Fle10}. Fig.~\ref{fig:qbo_bisongongvirgolf} shows the changes in $p$-mode frequency shifts due to the presence of the QBP signal for $\ell$=0, 1, 2 over solar cycle 23. 
GONG, BiSON, VIRGO and GOLF observations point to a modulation of the signal strength in phase with the solar cycle in agreement with previous findings \citep{Fle10,Bro11}. It also seems that the modulation affects more the $\ell$=1,2 modes compared to $\ell$=0. This is a further evidence of the higher sensitivity to low-mid latitudes of the sectoral modes. Therefore we might also argue that the observed QBP signal in the sectoral modes might come mainly from low-mid latitudes.
 If it is indeed the case, the signal at equatorial regions is of comparable strength (or even larger) to the QBP signal produced at high latitudes. This conclusion won't be correct at this stage.
The magnetic field, depending on its strength and inclination \citep{Sch06}, affects the size of the acoustic cavity, where the mode propagates. 
But as the acoustic cavity decreases when the spherical degree increases, therefore, to get information on the strength of the signal coming at different latitudes, we have to compare sectoral and azimuthal components of the same spherical degree.}
{Fig.~\ref{fig:qbo_gong_l12mcomp} shows the QBP signal over solar cycle 23 for the $m$-components of the dipole and quadrupole modes. 
The residuals for $\ell$=1, $m$=0 might indicate the presence of a QBP across July 1998 and soon after September 2006. 
If we look, instead, at $\ell$=1 $m$=1 component, we spot a nice-increasing trend in the signal strength of the QBP over the ascending and maximum phase of solar cycle 23. A further periodicity seems to occur across September 2006, and the same structure appears in GONG (integrated) and BiSON data.
 Looking at the zonal component of $\ell$=2, we find over the solar maxima the typical double peak structure and a further mid-term periodicity soon after September 2006 as in $\ell$=1, $m$=0. The residuals for the sectoral component shows a nice-increasing trend in the signal strength of the QBP over the ascending phase of solar cycle 23. A further 2-year modulation might have been occurred across September 2006 as found also in $\ell$=1, $m$=1. 
The azimuthal analysis confirms the modulation in the signal strength mainly occurring in the sectoral components of the modes and also that over the long extended minimum of solar cycle 23 the sources of the mid-term periodicity might be located at high latitudes.}
\subsection{QBP signal strength over high and low activity phase}
 {The zonal and sectoral components of the quadrupole mode show clear signatures of the QBP signal over periods coinciding with solar maxima. By comparing the peak-amplitude of the $m$-components over this activity phase, the signal seems to be of comparable strength at different latitudes. But as the strong magnetic fields are confined at low-mid latitudes, the comparison should be carried out over periods of low activity, when the magnetism of the Sun at low as well as high latitudes is dominated by weak fields. To this aim
 we might attempt to compare the strength of the possible signatures of the QBP signal we found over the years 2006-2009. For the quadrupole mode the strength is larger for the $m$=0 compared to $m$=2 component and the same feature occurs in the dipole mode. 
 As consequence this might be a further hint that the QBP signal at high latitudes is larger than the one at low-mid latitudes.}
\subsection{Investigating the significance of the QBP signatures over solar cycle}
{Evidences in favor of QBP have been found in all the modes. In order to investigate their significance, we applied the wavelet analysis developed by Torrence and Compo \citep{Tor98} to the spherical and azimuthal components of the modes. We show the results only for GONG observations, as the findings for the spherical degree analysis from the different datasets are in good agreement.}
\begin{figure*}
\mbox{\subfigure{\includegraphics[width=2.3in]{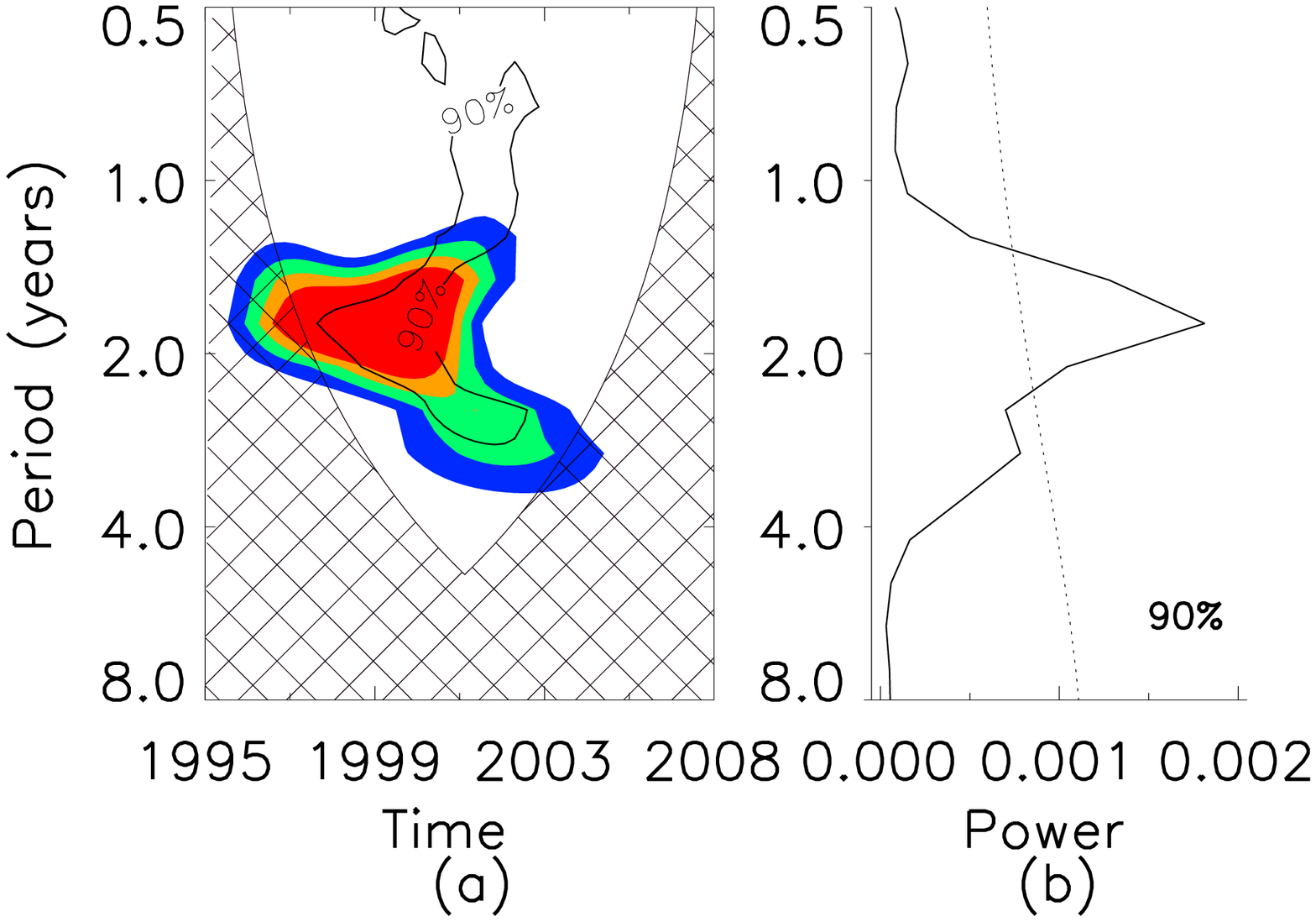}}
\subfigure{\includegraphics[width=2.3in]{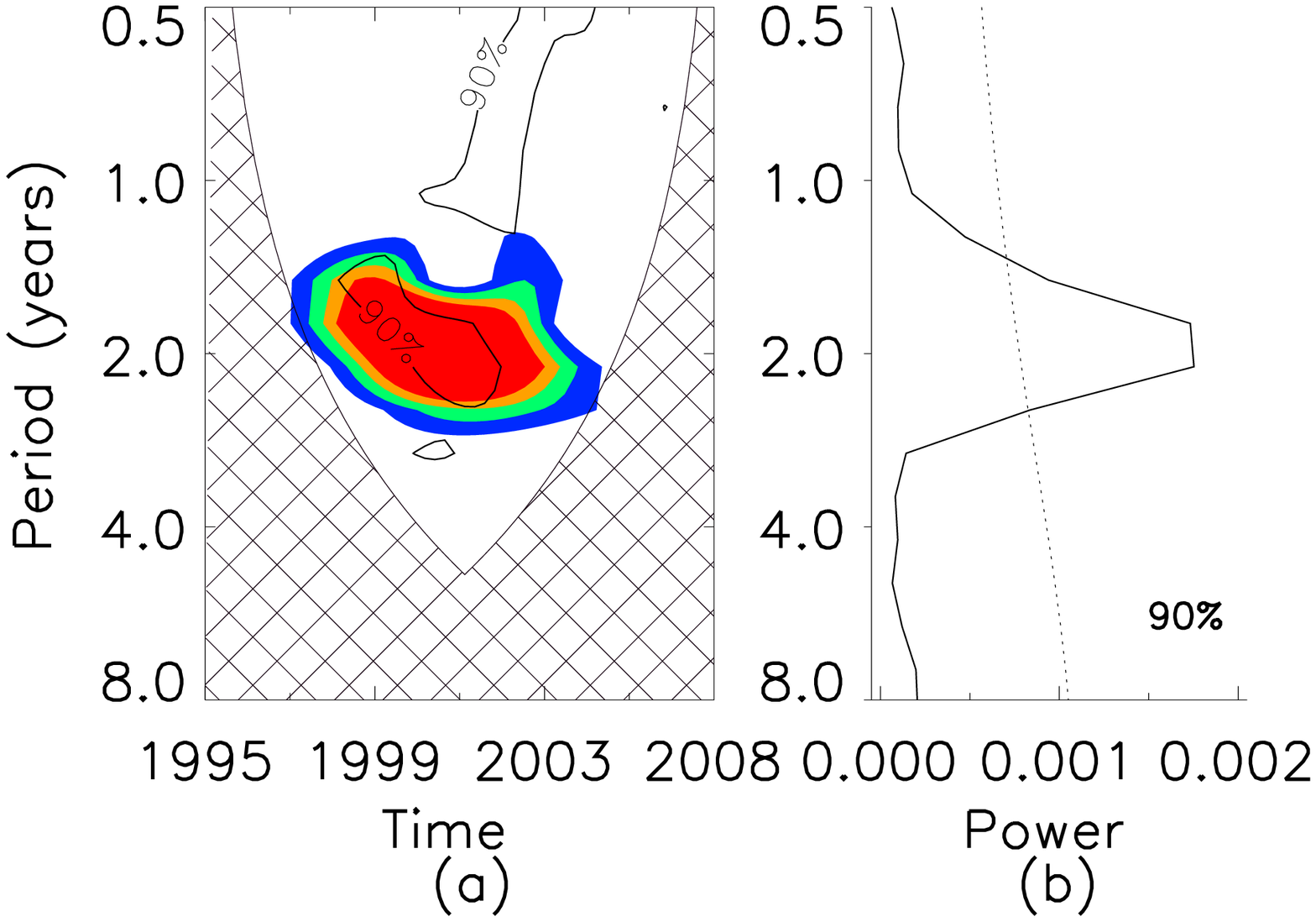}}}
\subfigure{\includegraphics[width=2.3in]{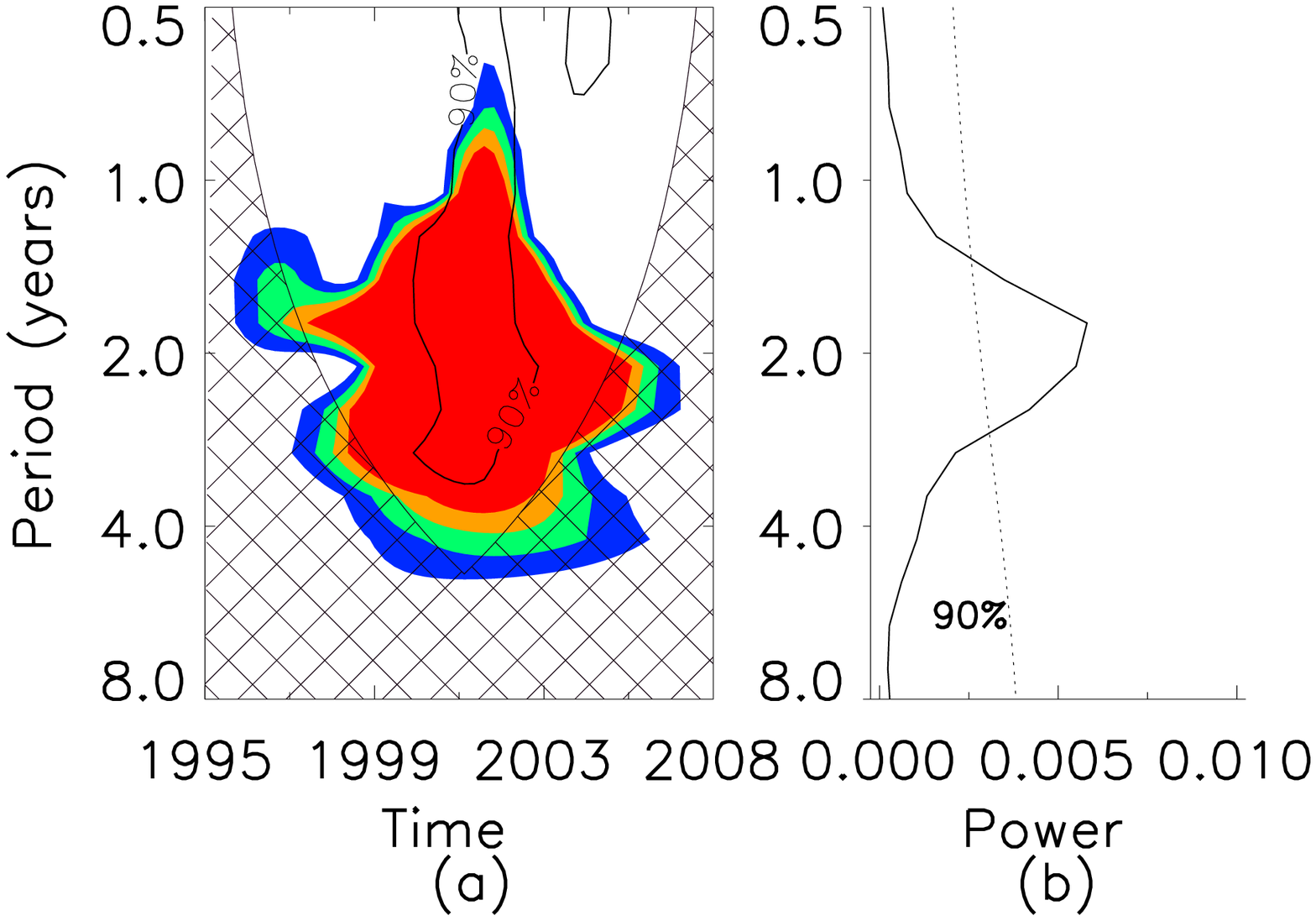}}
\caption{(a) The local wavelet power spectrum for $\ell$=0, 1, 2. White represent areas of little power, red those with the largest power. Solid rings identify areas at power of 0.001, 0.002,0.005 over time and only those that achieved 90$\%$ confidence level are labelled. (b)Global wavelet power spectrum. The dotted line are the 90$\%$ of confidence level.} 
\label{fig:wave_degree}
\end{figure*}
{Fig.~\ref{fig:wave_degree} shows the wavelet analysis carried out for each spherical degree. Within the cone of influence the local wavelet power spectrum identifies the QBP signal over the years 1998-2004 of solar cycle 23 in all low-degree modes. The significant periodicity at 90$\%$ of confidence level occurs at T=1.7 for $\ell$=0,2, while for $\ell$=1 we have two peaks above the 90$\%$ confidence level corresponding to T=1.7 and T=2.1.}
\begin{figure*}
\begin{center}
\mbox{\subfigure{\includegraphics[width=2.3in]{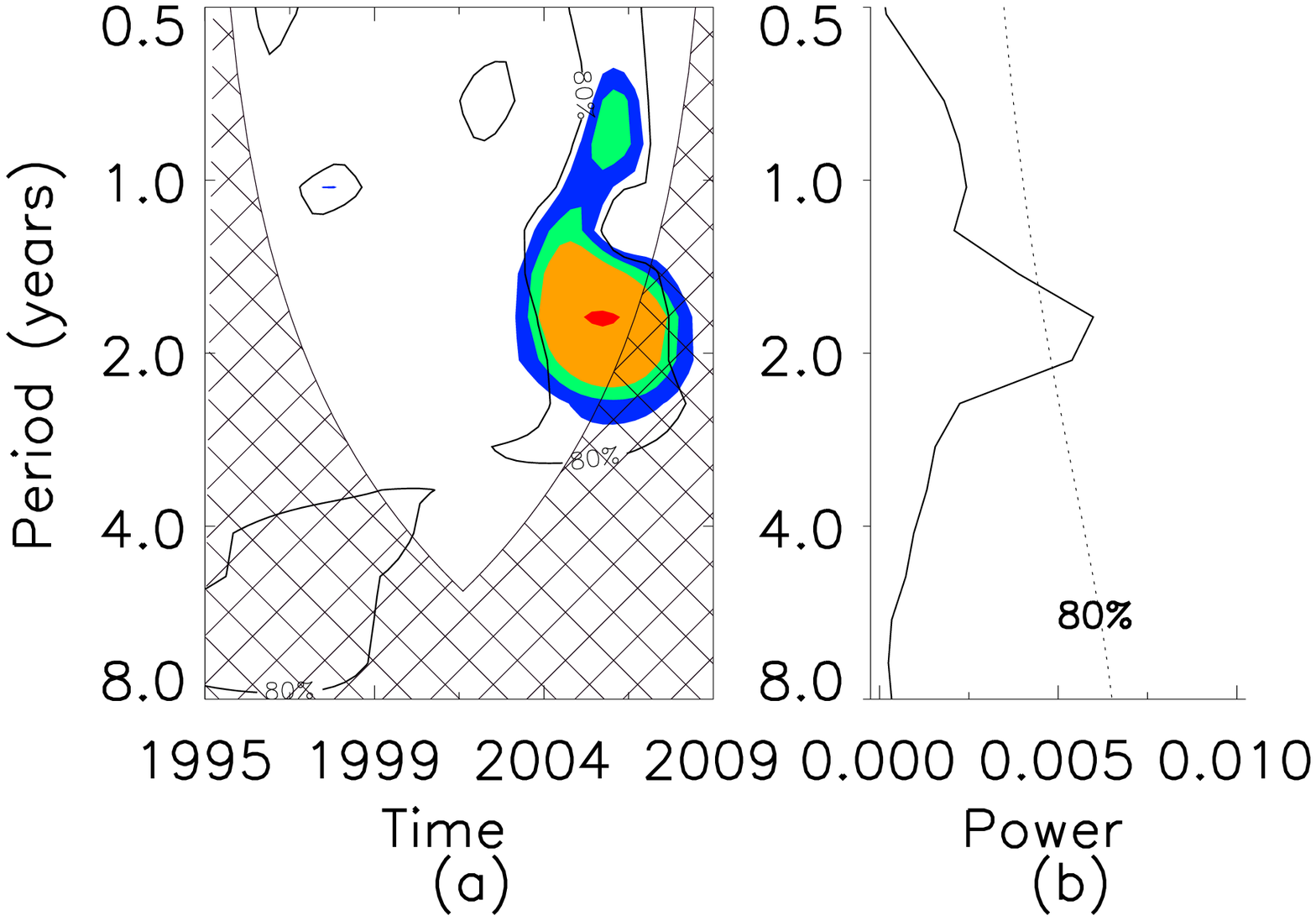}}
\subfigure{\includegraphics[width=2.3in]{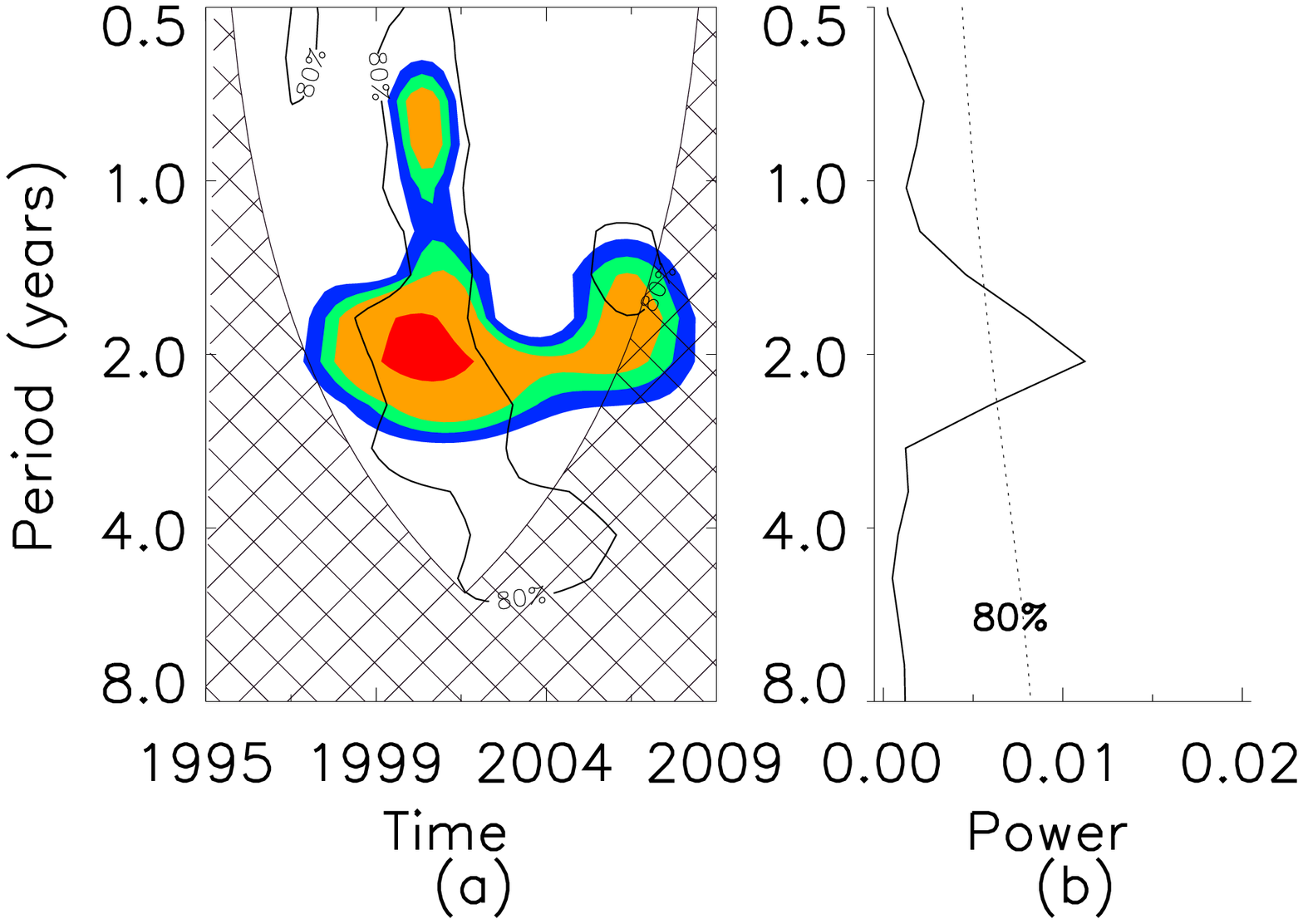}}}
\caption{The same as in Fig.5 for the $\ell$=1 $m$=0,1. Solid rings identify areas at power of 0.0050,0.0075,0.01 over time and only those that achieved 80$\%$ confidence level are labelled.} \label{fig:wave_azim1}
\end{center}
\end{figure*}
\begin{figure*}
\begin{center}
\mbox{\subfigure{\includegraphics[width=2.3in]{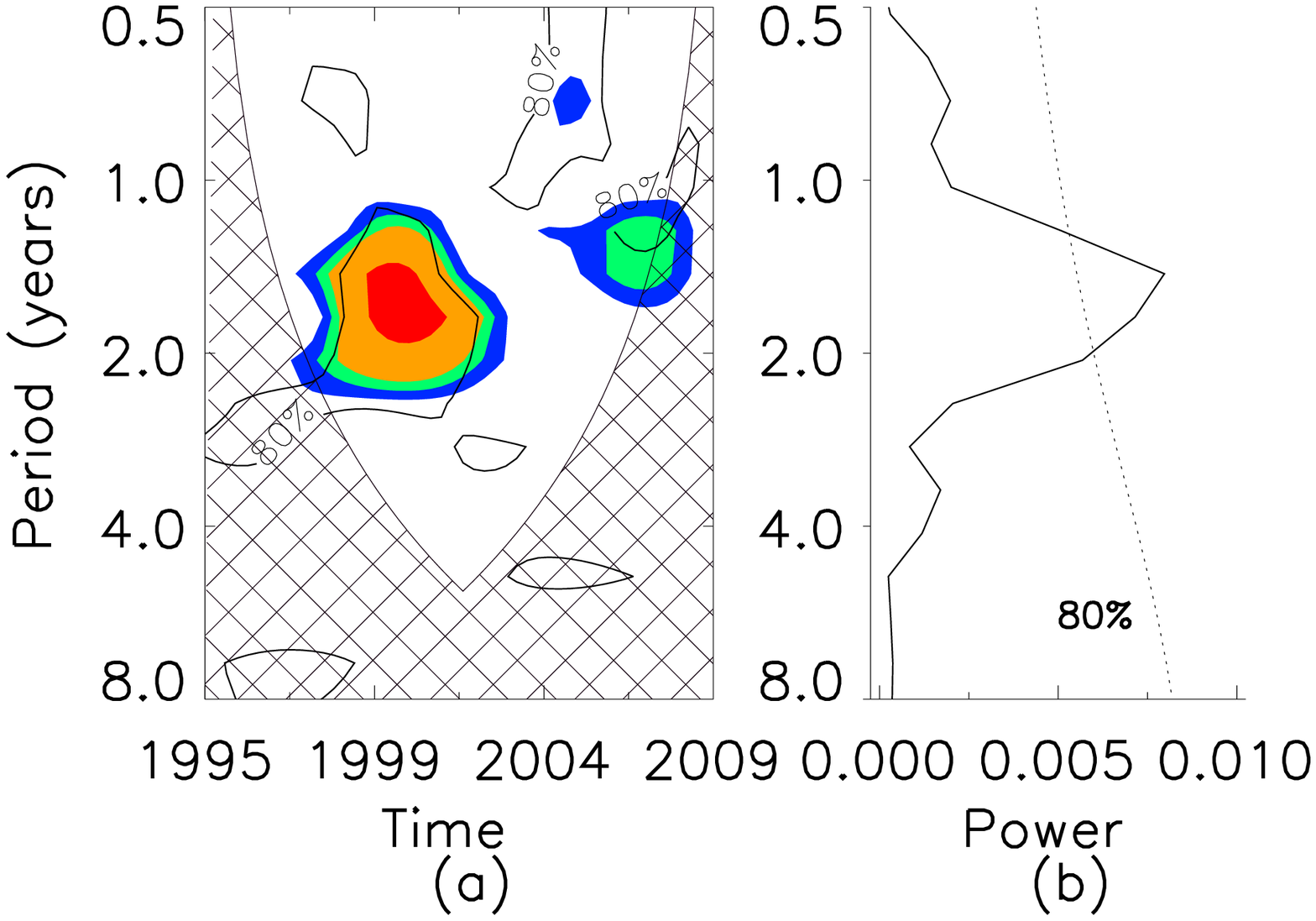}}
\subfigure{\includegraphics[width=2.3in]{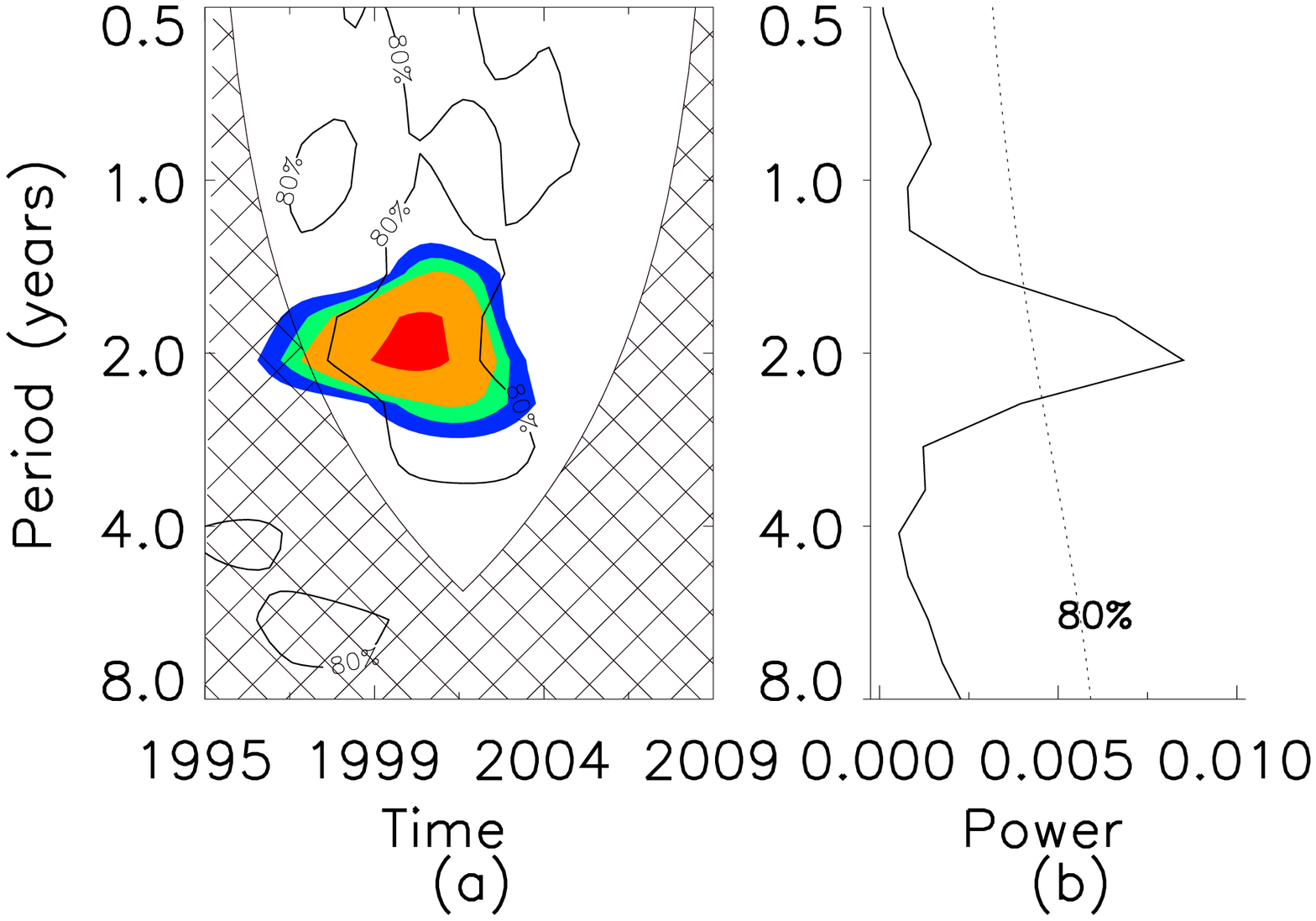}}}
\caption{The same as in Fig.6 for $\ell$=2 $m$=0,2.} \label{fig:wave_azim2}
\end{center}
\end{figure*}
{Fig.~\ref{fig:wave_azim1} -~\ref{fig:wave_azim2} shows the wavelet analysis for the azimuthal components of the modes. In this case we dropped down the significant level at 80$\%$ to show that there are signatures of QBP signal over the minimum activity phase in the zonal components of the modes. The significant periodicity at 80$\%$ of confidence level occurs at T=1.7 for $\ell$=1,$m$=0, T=2.1 for $\ell$=1,2 and $m$=1,2 and T=1.6 for $\ell$=2,$m$=0. It is interesting to note that the zonal modes show a slightly different periodicity from the one detected in the sectoral components. This might be the result of the greater sensitivity to higher latitudes of the zonal modes. To further investigate this point, it will be important to study the properties of the QBP signal in the azimuthal components of intermediate-high degree modes. There are evidences, in fact, showing that the periodicity from northern ($50^{0}\leq\theta\leq 90^{0}$) and equatorial ($- 45^{0}\leq\theta\leq 45^{0}$) regions differs from the one in the southern regions ($-50^{0}\leq\theta\leq -90^{0}$)\citep{Vec08}. 

 In summary the wavelet analysis confirms that a further periodicity of $\approx$~ 2-years is indeed present in the data at confidence level of 90$\%$.
Furthermore the findings seem also to point to a likely persistent nature of the QBP signal, since we found evidences that this mid-term periodicity is also present over periods of low activity phase.}
\section{$p$-mode frequency shift correlation coefficient with activity proxies over the 11-year cycle}
\subsection{Spherical degree dependence of the correlation coefficient}
The correlation analysis between $p$-mode frequency shift and surface activity measures might improve our understanding on the role played by strong and weak fields in the QBP signal. 
We, therefore,decided to investigate the degree of correlation between $p$-mode frequency shift and two activity proxies: MWSI (Mount Wilson Sunspot Index) and F10.7 index. The MWSI values are determined as a summation over pixels where the absolute value of the magnetic field strength is greater than 100 gauss.
 We have chosen MWSI because it is linked to the strong component of the toroidal fields generated by the global dynamo. These fields are mainly located at low-mid latitudes. F10.7 (RF) represents, instead, an integrated activity proxy and tells us the irradiance at 2.8~GHz at chromospheric heights coming mainly from 
the Quiet Sun background emission and also
Sunspots, Plages. This proxy therefore is a good measure of the weak (located at high latitudes) and strong magnetic fields (mainly at low-mid latitudes). 

We investigated the degree of correlation of $p$-mode frequency shift with MWSI and RF by splitting the whole cycle in different phases as follows:

- AP={\bf A}scending {\bf P}hase from 22 September 1996 until 26 June 1999

- MP= {\bf M}aximum {\bf P}hase 1999 June 27 to 2003 January 12

- DP= {\bf D}escending {\bf P}hase January 13 2003 26 until July 2007

- mP= {\bf m}inimum {\bf P}hase up to 2009.

\begin{figure}
\includegraphics[width=3.5in]{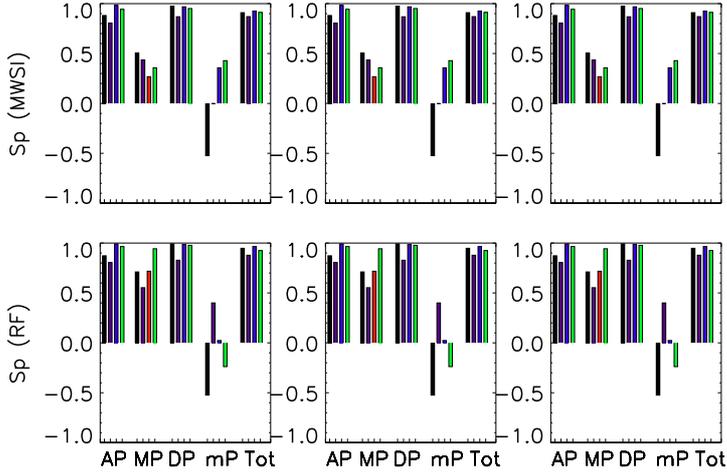}
\caption{The degree of correlation between $p$-mode frequency shifts of low degree modes ($\ell$=0,1,2 from left to right) and MWSI (top panel) and RF (bottom panel) over different phases of solar activity. The colors bar identify the observational programme, whose legend is the same as Fig.~\ref{fig:bisongongvirgolf}.} \label{fig:l012_bisongong}
\end{figure}

We used the Spearman's rank correlation in preference to the Pearson's correlation because the Pearson's rank correlation assesses the linear correlation between two sets of data. The Spearman's rank correlation is less specific. While some of the data sets we compare may be linearly related this is not definitely true in every case.
Fig.~\ref{fig:l012_bisongong} shows the correlation coefficient for $\ell$=0, 1, 2 mode frequency shifts with MWSI (upper panel) and RF (bottom panel) over different phases of solar cycle for BiSON, GONG, GOLF and VIRGO.
 We find that the degree of correlation of $p$-mode frequency shift with MWSI decreases over the maximum phase independently from $\ell$ value. If we look at RF we find higher correlation compared to MWSI all over different phases of the solar cycle.
These findings are in agreement with previous results \citep{Jai09} and they confirm that behind the mode frequency shift the active regions are not the only players. 
The other important feature is the {\it anticorrelation} over the minimum of solar cycle 23 for $\ell$=0, 2. In these modes there are traces of the QBP signal. 
This might  be the result of the different behavior of the surface activity measure over the long extended minimum. While MWSI went down to unprecedented level until 2009, Radio Flux shows already an upward trend in February 2008. This might explain the higher degree of anticorrelation with RF compared to MWSI. 
\subsection{Azimuthal dependence of the correlation coefficient}
 {It is possible to investigate the latitudinal dependence of the correlation coefficient by using the single $m$-components of the modes provided by GONG data.}
 
\begin{figure}
\begin{center}
\includegraphics[width=3.2in]{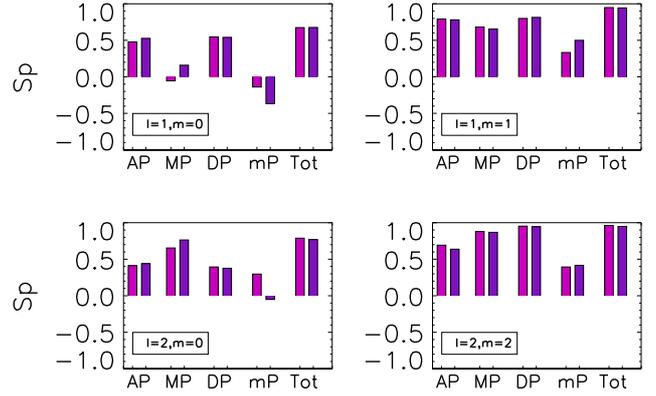}
\caption{The degree of correlation between $p$-mode frequency shifts of the azimuthal components of the modes and MWSI and RF. The colors bar identify MWSI (pink) and RF (purple).} \label{fig:l12mcomp_mwsi}
\end{center}
\end{figure}

 Fig.~\ref{fig:l12mcomp_mwsi} compares the degree of correlation between the zonal and sectoral components of dipole and quadrupole modes with MWSI and RF. The sectoral components seems to correlate better with both activity proxies over different phases of the solar cycle compared to the zonal modes. 
Previous works have shown that the central frequency of each multiplet is
best correlated with the global activity measure, while the individual m-component frequencies are more sensitive to the latitudinal distribution of activity \citep{How99,Ant01,How02}. Therefore the correlation analysis might suggest that the zonal modes are better correlated with resolved activity index that takes into account the latitudinal dependence of solar magnetic activity. To check upon this point, it will be important to further extend the correlation analysis to the azimuthal components of intermediate - high degree modes investigating the degree of correlation with resolved activity index.
\subsection{Conclusions}
{Evidences of QBP have been found in the individual low -$\ell$ degree modes as well as in their corresponding azimuthal $m$-components at all levels of solar activity. The strength of the QBP signal seems to be modulated by the 11-year envelope, but the effect is more significant at low-mid latitudes compared to higher ones. This modulation might explain the coupling between the strength of the main global dynamo and the appearance of the QBP mainly over periods coinciding with solar maxima. When the solar cycle is extremely weak, the magnetic flux processed by the main dynamo is not strong enough to bring up the mid-term periodicity signal in those surface activity indices related to the toroidal component of the main dynamo.

 The signatures of QBP have also been detected over the minimum phase of solar cycle 23 with 80$\%$ of confidence level. The likely persistent nature of the 2-year periodicity in the seismic analysis must be investigated, but probably shows a permanent and periodic phenomenon that stands in the sub-surface layers in addition to a pure presence of a magnetic field.
The occurrence of the QBP signal over periods of solar maxima as well as solar minima implies that the origin of the QBP is not coupled to the presence of active regions and therefore that the mechanism is different from the one that brings up to the surface the active regions. This finding is in agreement with previous results obtained by analyzing coronal holes data \citep{Vec08}.

 During periods of low activity the QBP signal from low-mid latitudes is almost absent, while it appears at higher latitudes. This finding might imply that the signal strength increases with latitudes and also that the sources of excitation might cover all range of latitudes. This feature might be explained within the formalism of a second dynamo mechanism situated near the sub-surface layer \citep{Ben98a,Ben98b}.  
 
 The seismic investigation brings new insight on the features of solar magnetic activity that might be useful to improve dynamo models. To analyze the signature of the QBP further investigation of the sectoral and zonal components of higher degree modes is strongly desired.  
We should also carry out a more detailed investigation on the latitudinal dependence of the signal strength over solar cycle. The findings could be extremely useful to fully identify the physical mechanism behind it, as at this early stage we cannot completely rule out the magnetic Rossby instability or other dynamo models that already take into account a further periodic component \citep{Flu04}. 
 Moreover we would be able to check whether or not the periodicity differs in the two hemisphere as already pointed out by \cite{Vec08}. 

The end of the solar cycle 23, characterized by its long extended minimum and the ongoing solar cycle is a current subject of research. 
  These new solar cycle features has re-opened interests and debates on the dynamo mechanism that rules solar magnetic activity. 
\acknowledgements
This work has been supported by the Swiss National Funding 200020-120114. This work utilizes data obtained by the Global Oscillation Network
Group (GONG) program, managed by the National Solar Observatory, which
is operated by AURA, Inc. under a cooperative agreement with the
National Science Foundation. The data were acquired by instruments
operated by the Big Bear Solar Observatory, High Altitude Observatory,
Learmonth Solar Observatory, Udaipur Solar Observatory, Instituto de
Astrof\'{\i}sica de Canarias, and Cerro Tololo Interamerican
Observatory. Wavelet software was provided by C. Torrence and G. Compo, and is available at URL: http://atoc.colorado.edu/research/wavelets/. 
 We thank members of the BiSON team, both past and present, for their technical and analytical support. We also thank Y.~Elsworth, W.~J.~Chaplin and A.-M. Broomahall for providing the data and useful comments to improve the manuscript. DS acknowledges the financial support from the Centre National d'Etudes 
Spatiales (CNES).
RS also is grateful to S.Turck-Chieze and CEA/IRFU for providing the facilities required to continue her work. We also thank Dr.T.Zaqarashvili for useful discussion and suggestions on the interpretation of the results.
\bibliographystyle{apj}
\bibliography{biblio1}

\begin{table*}
\begin{center}
\begin{tabular}{c c c c c c c c c c c }\hline\hline
{\bf Activity Index} & {\bf Mode of oscillation} & {\bf Data Source} &\multicolumn{2}{c}{\bf AP}&\multicolumn{2}{c}{\bf MP}&\multicolumn{2}{c}{\bf DP}&\multicolumn{2}{c}{\bf mP}\\ \hline
MWSI&$\ell$=0&BiSON& 0.88&$3^{-5}$&0.50&0.06&0.98&$1^{-12}$&-0.52&0.18\\ 
&&GONG& 0.80&$2^{-3}$&0.43&0.09&0.87&$1^{-5}$&0&0.1\\ 
&&VIRGO& 0.94&$4^{-7}$&0.36&0.2&0.95&$4^{-10}$&0.42&0.3\\ 
&&GOLF&-0.98&$4^{-10}$&0.27&0.3&0.97&$1^{-11}$&0.36&0.4\\ \hline

&$\ell$=1&BiSON&0.91&$4^{-6}$&0.52&0.04&0.94&$2^{-9}$&0.26&0.5\\ 
&&GONG&0.97&$1^{-7}$&0.63&$8^{-3}$&0.82&$1^{-5}$&-0.20&0.8\\ 
&&VIRGO&0.92&$3^{-6}$&0.40&0.13&0.94&$3^{-9}$&0.67&0.08\\ 
&&GOLF&0.92&$4^{-6}$&0.2&0.5&0.96&$2^{-11}$&0.52&0.18\\ \hline

&$\ell$=2&BiSON&0.92&$4^{-6}$&0.52&0.05&0.94&$2^{-9}$&0.26&0.5\\ 
&&GONG&0.87&$2^{-4}$&0.60&0.01&0.95&$2^{-10}$&-0.2&0.8\\ 
&&VIRGO&0.77&$1^{-3}$&0.46&0.08&0.92&$3^{-8}$&0.19&0.65\\ 
&&GOLF&0.90&$1^{-5}$&0.48&0.07&0.90&$1^{-7}$&0.36&0.38\\ \hline

RF&$\ell$=0&BiSON&0.87&$5^{-5}$&0.71&$3^{-3}$&0.99&$2^{-16}$&-0.52&0.2\\ 
&&GONG&0.80&$2^{-3}$&0.55&0.02&0.83&$1^{-5}$&0.40&0.6\\ 
&&VIRGO&0.96&$2^{-8}$&0.94&$1^{-7}$&0.97&$7^{-13}$&-0.24&0.57\\ 
&&GOLF&0.99&$6^{-12}$&0.72&$2^{-3}$&0.98&$1^{-14}$&0.023&0.9\\ \hline

&$\ell$=1&BiSON&0.97&$4^{-9}$&0.68&$5^{-3}$&0.95&$2^{-10}$&0.88&$4^{-3}$\\ 
&&GONG&0.97&$1^{-7}$&0.74&$9^{-4}$&0.80&$5^{-5}$&0.93&$4^{-11}$\\ 
&&VIRGO&0.95&$2^{-6}$&0.58&0.02&0.93&$4^{-11}$&-0.14&0.7\\ 
&&GOLF&0.96&$9^{-8}$&0.67&$6^{-4}$&0.98&$1^{-13}$&-0.05&0.9\\ \hline

&$\ell$=2&BiSON&0.97&$4^{-9}$&0.89&$7^{-6}$&0.91&$7^{-8}$&-0.86&$6^{-3}$\\ 
&&GONG&0.87&$2^{-4}$&0.49&0.06&0.94&$1^{-9}$&-0.40&0.6\\
&&VIRGO&0.77&$1^{-3}$&0.83&$1^{-4}$&0.93&$8^{-9}$&-0.43&0.3\\ 
&&GOLF&0.91&$5^{-6}$&0.89&$7^{-6}$&0.93&$1^{-8}$&-0.78&0.02\\ \hline
\end{tabular}
\caption{The Spearmann correlation coefficient and significance between Activity Indeces and low degree $p$-mode frequency shift over different phases of solar activity.}
\label{table:coeff_mwsirf_bisongong_l0}
\end{center}
\end{table*} 
 
\begin{table*}
\begin{center}
\begin{tabular}{c c c c c c c c c c}\hline\hline
{\bf Activity Index}&{\bf Mode of oscillation}&\multicolumn{2}{c}{\bf AP}&\multicolumn{2}{c}{\bf MP}&\multicolumn{2}{c}{\bf DP}&\multicolumn{2}{c}{\bf mP}\\ \hline
MWSI &$\ell$=1,$m$=0&0.48&0.1&-0.05&0.8&0.55&0.03&-0.14&0.7\\ 
&$\ell$=1,$m$=1&0.79&$1^{-3}$&0.68&0.01&0.80&$1^{-4}$&0.33&0.34\\ 
&$\ell$=2,$m$=0& 0.41&0.16&0.65&0.01&0.39&0.13&0.30&0.13\\ 
&$\ell$=2,$m$=2&0.69&$9^{-3}$&0.88&$7^{-5}$&0.95&$1^{-8}$&0.39&0.3\\ \hline

RF&$\ell$=1, $m$=0& 0.52&0.06&0.16&0.6&0.54&0.03&-0.36&0.3\\ 
&$\ell$=1, $m$=1& 0.78&$1^{-3}$&0.65&0.01&0.81&$1^{-4}$&0.50&0.2\\ 
&$\ell$=2, $m$=0& 0.44&0.13&0.69&0.02&0.38&0.15&-0.57&0.89\\ 
&$\ell$=2, $m$=2&0.64&0.02&0.87&$1^{-4}$&0.94&$4^{-8}$&0.95&0.3\\ \hline
\end{tabular}
\caption{The same as Table 1 but for the azimuthal component of low degree modes.}
\label{table:coeff_rf_mcomp}
\end{center}
\end{table*}

\end{document}